\documentclass[letterpaper,11pt,leqno]{article}
\usepackage{graphicx}
\usepackage[margin=1in]{geometry}
\usepackage{paper}
\hypersetup{pdftitle={Central Bank Communication with Public:  Bank of England and Twitter (X)}}

\newcommand{\bib}{bibliography.bib}

\newcommand{\pdf}{figures.pdf}

\title{
  {\fontsize{23pt}{26pt}\selectfont 
  Central Bank Communication \\ with Public:  \\  Bank of England and Twitter (X)}\thanks{The views expressed are those of the authors and do not necessarily reflect the views of any institution.}%
}
\author{
  \sc Fatih Kansoy%
  \thanks{Contact: \href{mailto:fatih.kansoy@economics.ox.ac.uk}{fatih.kansoy@economics.ox.ac.uk}} \\
   University of Oxford
  \and
  \sc Joel Mundy %
    \thanks{Contact: \href{mailto:joel.mundy@bankofengland.co.uk}{joel.mundy@bankofengland.co.uk}} \\
     Bank of England
}
\date{May 2025}
\begin{document}
\maketitle

\available{https://www.fatih.ai/boe.pdf}
\thispagestyle{plain}

{\setstretch{1}
\begin{abstract}

\noindent  Central banks increasingly use social media to communicate beyond financial markets, yet evidence on public engagement effectiveness remains limited. Despite 113 central banks joining Twitter between 2008-2018, we lack understanding of what drives audience interaction with their content. To examine engagement determinants, we analyzed 3.13 million tweets mentioning the Bank of England (2007-2022), including 9,810 official posts. We investigate posting patterns, measure engagement elasticity, and identify content characteristics predicting higher interaction.
The Bank's posting schedule misaligns with peak audience engagement times, with evening hours generating highest interaction despite minimal posting. Cultural content (Alan Turing £50 note) achieved 1,300 times higher engagement than routine policy communications. Engagement elasticity averaged 1.095 with substantial volatility during events like Brexit, contrasting with the Federal Reserve's stability. Media content dramatically increased engagement: videos (1,700\%), photos (126\%), while monetary policy announcements and readability significantly enhanced all metrics. Content quality and timing matter more than posting frequency for effective central bank communication. These findings suggest central banks should prioritize accessible, media-rich content during high-attention periods rather than increasing volume, with implications for digital communication strategies in fulfilling public transparency mandates.

\medskip


\medskip

\end{abstract}
}

\clearpage
\pagebreak 
\newpage


\onehalfspacing

\section{Introduction}

Central banks have undergone a profound transformation in their communication strategies over recent decades, transitioning from a culture of opacity to one characterized by transparency and active public engagement \citep{blinder2008central, issing2020longjourney}. This evolution reflects the growing recognition that communication itself constitutes a vital monetary policy instrument \citep{coibion2022monetarypolicycommunications}. Historically, central banking was shrouded in secrecy – epitomised by former Bank of England Governor Montagu Norman's dictum "\textit{never explain, never excuse}" – where policymakers deliberately withheld information from the public \citep{bernanke2007federal}. This approach continued through the era of Alan Greenspan, who as Federal Reserve Chairman employed what became known as "Fedspeak" – a "language of purposeful obfuscation" characterized by deliberately ambiguous statements. Greenspan himself acknowledged this strategy, once remarking to Congress, "\textit{If I seem unduly clear to you, you must have misunderstood what I said}" demonstrating how central bankers actively avoided transparency \citep{geraats2018mystique}.

The paradigm shift toward openness accelerated under Ben Bernanke's Federal Reserve chairmanship, who famously characterised monetary policy as "\textit{98 percent talk and 2 percent action}" \citep{bernanke2015inaugurating}. Modern theoretical frameworks, including Barro-Gordon and New Keynesian models, demonstrate that credible, clear communication can bolster central bank legitimacy, reduce uncertainty in financial markets, and strengthen monetary policy transmission \citep{gorodnichenko2021voice,lamla2019central,coibion2022monetarypolicycommunications}. By openly communicating policy objectives and decisions, central banks create informative news that complements their actions while reducing noise and uncertainty in public interpretation of policy. This approach enhances policy predictability, helps anchor expectations, and serves a democratic accountability function by legitimising independent central banks through transparent dialogue about policy rationales \citep{bernanke2004monetary}.

The 2008 global financial crisis marked another watershed moment, as unconventional monetary policies like forward guidance and quantitative easing further elevated the importance of communication as a policy tool. Central banks not only increased the frequency and scope of their communications but also expanded their target audience beyond financial experts to include the general public \citep{assenmacher2021clear,masciandaro2024central_joes}. Despite this expanded outreach, empirical evidence suggests limited public engagement with traditional central bank communications. \cite{kumar2015inflationtargetingnz} found minimal readership of Reserve Bank communications in New Zealand, while \cite{vandercruijsen2015publicknowecb} documented poor knowledge of the European Central Bank's objectives among Dutch households, along with little interest in becoming better informed. This apparent disconnect presents a significant challenge as central banks increasingly seek to communicate directly with broader audiences \citep{haldane2018little,bholat2019enhancing}.

The rise of social media platforms, particularly Twitter (recently rebranded as X), has provided central banks with novel channels to disseminate information widely and interactively. This shift represents a substantial evolution in how central banks conceptualize their communication strategies and target audiences, offering potential solutions to the engagement gap identified in previous research. A growing body of evidence suggests that effective communication through these channels can shape not only market expectations of future policy but also influence household inflation expectations \citep{binder2017fedspeak,binder2023central,coibion2022monetarypolicycommunications,kryvtsov2021communicationworks}. For example, \cite{angelico2022can} demonstrate that real-time social media dialogue can capture shifts in inflation sentiment, while \cite{ehrmann2022central} highlight the connection between transparency, accountability, and central banks' approach to non-expert audiences.

\paragraph{Adoption, Content, and Engagement} Central banks worldwide have rapidly adopted social media, especially Twitter, (e.g., \cite{kyriakopoulou2019central} documented that 113 central banks joined Twitter between 2008 and 2018) to reach broader audiences. \cite{blinder2024communicationgeneralpublic} find that Twitter is the most popular platform for central banks: nearly all of a sample of 75 central banks (including all major ones) now maintain official Twitter accounts\footnote{Notable exceptions are rare; for example, the People's Bank of China doesn't use X/Twitter but engages the public via Weibo, which is China's equivalent of X in China}. This uptake spans both advanced and emerging economies. In fact, emerging market central banks are among the most active Twitter users - central banks in Latin America (e.g. El Salvador, Mexico, Ecuador, Argentina) rank in the top 10 by tweet frequency, demonstrating a particularly strong social media presence \citep{blinder2024communicationgeneralpublic}. The Bank of England was a pioneer in central bank adoption of social media, launching its Twitter account (@bankofengland) in January 2009, making it the one of the first European central bank to establish a presence on the platform \citep{masciandaro2024central_joes}. Initially, the account was used sparingly, primarily to signpost users to information published on the official website. For example, while the Bank of England joined Twitter in January 2009, it did not tweet until July 2011\footnote{Here is the first tweet from the Bank of England's Twitter account: \url{https://x.com/bankofengland/status/86720699535851520}}. Over time, the Bank's approach evolved to include more direct engagement with the public.

However, central banks vary widely in how they use Twitter. In a multi-year survey, \cite{korhonen2019masteringtwitter}   examined the Twitter activity of 40 European central banks and financial supervisors, finding large disparities in tweeting frequency and content across institutions. \cite{masciandaro2024central_joes} show that announcements about new banknotes or commemorative coins often generate disproportionate engagement. Yet tweets explicitly on monetary policy can also garner widespread attention if they clarify policy shifts or respond quickly to breaking developments \citep{kyriakopoulou2019central}. Analyses of content typically reveal a blend of monetary policy statements, financial stability bulletins, educational infographics, and community outreach messages \citep{haldane2018communicationspublic,gorodnichenko2021voice}. For instance, European central banks (ECB) employ “layered communication,” mixing expert-level detail (interest-rate decisions, forward guidance) with plainer language for the public \citep{ehrmann2022socialmedia}. As underlined by \cite{lamla2022gentlemantweet} the Bank of England stands out for its efforts to use clear and accessible language on Twitter. Unlike the Fed and ECB, which often tweet links to press releases or speeches, the BoE's Twitter team presents information in a simplified form, using 'simple words and concise infographics'. The BoE was one of the first central banks to actively incorporate such plain-language content in its social media posts. For example, after each Monetary Policy Committee meetings, the BoE tweets key decisions with graphical summaries and minimal jargon, aiming to make policy news digestible for non-experts. Research suggests this strategy can indeed make a difference: \cite{haldane2018communicationspublic} found that providing information in an easy-to-understand format (versus a technical release) led to stronger belief updates by the public, lending support to the BoE's approach.

A key question in recent research is what content central banks share on social media and how the public engages with it. Although few, the literature reveals several patterns in the types of posts and their reception. When central banks do tweet about core monetary policy decisions or new banknotes or coins and currency design, those posts tend to see bumps in engagement. For instance, \cite{gorodnichenko2024socialmedia} found the Fed's tweets about monetary policy and economic conditions drew significantly more user interactions than tweets on peripheral topics. In their sample, tweets referencing \textit{inflation}, \textit{unemployment}, or \textit{financial stability} issues elicited more retweets, indicating higher interest when central banks address topics clearly within their mandate.

In addition the topics, communication style and tone are important for public engagement. How a message is conveyed can determine its reach. \cite{ehrmann2022socialmedia} highlight that on Twitter tone matters - tweets that used simpler, more neutral language tended to just relay information, whereas tweets containing strong opinions or dramatic wording (often by non-official accounts commenting on the ECB) were much more likely to be shared widely. Another aspect of style is clarity. \cite{korhonen2024microblogging} focus on the readability of ECB communications (including tweets, speeches, and press conference statements) and its impact on public engagement. They conclude that greater clarity leads to stronger engagement - when the ECB's language is easy to understand, its social media posts garner more likes, retweets, and replies. This implies that clarity itself, not just the importance of the news, drives people to interact. 

Central banks measure success on social media partly by engagement metrics (e.g. retweets, likes, follower growth). For example, \cite{masciandaro2024central_joes} shows that, on average, only ~2.5\% of G20 central banks' tweets were direct replies to other users. Thus, while central banks are active online, they are generally not using Twitter as a forum for dialogue but rather as a one-way publishing platform. Public engagement is mostly measured in how people react to central bank posts, not two-way communication. \cite{haldane2018little} underline the importance of two-way communication (effective communication), particularly for institutions like central banks, needs to evolve beyond traditional one-way pronouncements ("communication means mouths") towards genuine two-way conversation ("conversation means ears as much as mouths"). This shift necessitates less focus on simply conveying information and more on engagement, which includes actively listening to and understanding the public ("understanding the public")  to build trust and legitimacy. Reaching a wider, more diverse audience beyond just experts and markets requires adapting language to be simpler and more relatable, using personalised and localised messaging, and employing new methods to connect with previously unreached segments of society. Ultimately, fostering public understanding and trust involves more listening and conversation, and potentially less unilateral "action". In this regard, we can say that the Bank of England stands out as one of the most active central banks on Twitter, using the platform not only to disseminate information but also to engage directly with individual users \footnote{\begin{minipage}{\linewidth}
    For example, the following exchange, which took place on X (Twitter at that time) on February 3, 2022, captures a dialogue between Brendan Dempsey (@brendempsey) (an ordinary Twitter user) and the Bank of England (@bankofengland) concerning the Bank's decision to raise interest rates to 0.5\%. 
        
    At 12:54 UTC on February 3, 2022, Brendan Dempsey initiated the conversation by posing a question to the Bank of England: \textit{"Can you explain how increasing interest rates helps do this? I'm assuming there's a fiscal logic, but on the face of it, your response to mounting food, energy, insurance, water, and petrol prices is to also hike mortgage repayments, which will also hit renters."} 
        
    The Bank of England responded at 15:17 UTC on the same day, stating: \textit{"We understand that this is a difficult time for people across the country. We also understand that monetary policy has little impact on international energy prices. However, the best contribution we can make to the UK economy is to stabilise inflation in the medium term. To do this, we have determined that a rise in Bank Rate is required. I would encourage you to view the Monetary Policy Report (MPR) and Press Conference for further information: [https://t.co/4Ia7gRGWgR]."} 
        
    At 15:29 UTC, Brendan Dempsey replied, expressing appreciation but seeking further clarification: \textit{"Thanks for replying. I understand inflation needs to be curbed, but what I asked—and your MPR doesn't explain—is how a rise in the Bank Rate does that. To combat rising inflation and an increased cost of living, you're further increasing the cost of living. How does that help?"} 
        
    The Bank of England then provided a follow-up response, also on February 3, 2022, explaining: \textit{"Inflation in essence shows an imbalance between supply and demand. Monetary policy is a demand-side policy, and as such, a rise in Bank Rate reduces demand."} see full conversation at \url{https://x.com/brendempsey/status/1489220809361133574}
    \end{minipage}}

\paragraph{Effects on Market Expectations and Public Understanding} One of the ultimate goals of central bank communication is to shape expectations -  whether it be financial market expectations for interest rates or the public's expectations for inflation and the economy. A crucial question is how social media communication feeds into this goal. The evidence so far is mixed, indicating some benefits of new channels as well as limitations. \cite{lamla2022gentlemantweet} provides a nuanced look at this issue. Their study on Bank of England communications found that while policy announcements didn't immediately change average public expectations, individuals who received the news, often via channels like Twitter, showed better-informed economic perceptions. Twitter helped spread information wider, increasing the number of informed people. However, those following the news on Twitter tended to overestimate inflation and interest rates, being overly confident despite their errors. This suggests that although social media expands reach, traditional media might still be better for nuanced understanding, highlighting the challenge of ensuring messages are not just heard, but correctly understood.

Financial markets represent another central bank Twitter audience, though traders typically rely on faster channels for real-time reactions. \cite{hansen201815} showed that specific wording in Bank of England communications measurably affected financial markets, with changes in language around forecasts moving investors' inflation and interest rate expectations. This indicates markets respond to both policy decisions and their communication style. While Twitter isn't traders' primary information source, it reinforces market messages. Relatedly, \cite{gorodnichenko2024socialmedia} found positive sentiment in central bank tweets correlated with higher public inflation expectations during certain periods—suggesting social media tone can influence sentiment marginally, as when optimistic Fed tweets during low-rate environments aligned public expectations with Fed goals.

Beyond markets, research examines the penetration of central bank communication into public awareness. Before social media, studies found widespread ignorance of central bank actions and inflation targets among households. \cite{binder2017fedspeak} documented that many U.S. consumers neither knew the Fed's inflation goal nor accurately perceived recent inflation, despite post-2008 communication expansions—a disconnect she termed "Fed speak on Main Street," highlighting that increased communication doesn't automatically improve public understanding. Social media potentially bridges this gap by enabling direct, unfiltered communication with the public. \cite{ehrmann2022socialmedia} analysis of ECB Twitter traffic offers encouraging evidence: major announcements (like Draghi's "whatever it takes" speech) generated significant Twitter activity that evolved from initial emotional reactions toward more factual discussions centered on official ECB publications, demonstrating how accurate information gradually crowds out misinformation when central bank messaging reaches sufficient audiences with clarity. Experimental research further confirms that simplified policy statements consistently improve public understanding and expectation accuracy compared to traditional communications, underscoring the critical importance of clarity and targeted engagement strategies when communicating with the general public.

In summary, the literature suggests that communication via Twitter has expanded the audience for central bank messages and can shape expectations at the margin, but its effectiveness hinges on how messages are crafted and received. Simply having a Twitter account is not a panacea for public understanding – content is king, and central banks must still battle confusion and misinterpretation among non-expert audiences.

\paragraph{Transparency, Trust, and Accountability} Using social media for outreach is part of central banking's broader transparency push, valued for both economic benefits and reinforcing institutional legitimacy. Many publications link improved communication with higher public trust. For example, in a 2018 speech, the Bank of Canada Governor Stephen Poloz emphasised that transparency helps build trust with the public and markets, making policy more effective \citep{poloz2018transparency}. 

One challenge in transparency is addressing multiple audiences. "Layered" communication addresses multiple audiences simultaneously. The Bank of England's communications reform tackled this by producing different tiers: in-depth releases for experts and simplified summaries with infographics for the public. Testing around 2018 showed improved public recall and understanding, leading to implementations like "Inflation in 5 minutes" and increased community outreach. Other central banks have adopted similar approaches, recognising that transparency requires not just information disclosure but engaging, comprehensible communication\citep{hansen2018bank} .

Social media offers central banks direct, unmediated  communication channels to reach the public. During crises, these platforms enable real-time clarifications and counter misinformation. It is clear that in times of crisis, central bank tweets can correct false information online. While \cite{blinder2024communicationgeneralpublic} found modest early results in reaching non-experts, "glimmers of hope" emerge when communication becomes two-way, as with the ECB's public forums and the Fed's town halls. However, risks include message oversimplification due to character limits and potential reputational damage from politicised discussions.

The evolution "from silence to Twitter" is evident as central banks now routinely tweet updates once confined to press releases. Research indicates this evolution can enhance transparency and public understanding when executed with clarity, consistency, and audience adaptation. Current literature largely agrees that social media engagement, while challenging, is worthwhile for fostering informed, trustful dialogue between central banks and society.

This paper contributes to the literature by examining whether and how the public engages with the Bank of England on Twitter. Specifically, we make three contributions. First, we provide a comprehensive descriptive analysis of the Bank of England's communication strategy through Twitter and documenting the evolution of this approach over time. Second, by utilising the  modern natural language processing techniques, we conduct a detailed analysis of public engagement patterns with the Bank of England on Twitter, identifying the types of content that generate the most significant responses. Third, we empirically investigate the determinants of engagement with Bank of England tweets, analysing factors that predict higher levels of public interaction.

Our research is situated within the broader literature on central bank transparency and digital communication strategies in the post-crisis era, extending that literature with new empirical evidence from a real-world setting. Our study leverages an extensive dataset that combines both sides of the communication exchange: all tweets posted by or referencing the Bank of England's official Twitter account. By linking the content of the Bank's communications with the social media reactions they provoke, we investigate which types of messages resonate most with the public.

This research offers several contributions to the existing literature. First, it provides novel evidence on public engagement with central bank content in a natural setting, as opposed to the controlled experiments or survey-based approaches that dominate the existing literature on non-expert audiences (e.g., \cite{coibion2020inflationpolicytool}; \cite{binder2017fedspeak}). Second, our focus on the Bank of England—a pioneer in adopting plain-language communication and social media outreach—allows us to evaluate the real-world efficacy of a leading central bank's digital engagement strategy. Third, by analyzing a comprehensive dataset of tweets about the central bank (not just the Bank's own posts), we shed light on the broader public discourse surrounding central bank communications over time.

The findings of this study aim to enrich our understanding of central bank communication in the digital age and to inform ongoing debates on transparency and engagement. In particular, identifying which types of content and messaging strategies garner the most public interaction can provide practical insights for central banks seeking to improve their outreach. Ultimately, by examining the Bank of England's experience with Twitter communication, we contribute new evidence on the opportunities and limitations of social media as a tool for enhancing the public's connection with monetary policy.

The remainder of this paper is organized as follows.  Sectio-\ref{sec:data} describes our data collection methodology and presents descriptive analyses of Bank of England communications and public responses. Section-\ref{sec:methodology} outlines our empirical methodology for analysing engagement determinants. Section-\ref{sec:results} presents our findings, and Section-\ref{sec:discussion} discusses implications and concludes.

\section{Data Collection and Description}\label{sec:data}

To analyse the Bank of England's (BoE) Twitter communication strategy and public engagement, we constructed a comprehensive dataset spanning from Twitter's inception in 2006 through July 2022. Data collection employed the SNScrape Python module and Selenium for large-scale Twitter data retrieval. This methodological approach was particularly valuable as it predated Twitter's 2023 rebranding to "X" and subsequent implementation of restrictive data access policies that now require enterprise-level fees monthly for comparable data collection capabilities. Second, over the period of study, patterns of usage of study followed consistent trends. However, once Twitter's ownership changed, actions by both owners and users led to changes in how Twitter was used. This creates challenges in studying the communication of central banks on Twitter over this period. We avoid this challenge by using data prior to this period.

Our data collection process encompassed two distinct components:

\begin{enumerate}
    \item \textbf{Official BoE communications}: All tweets from the Bank of England's verified Twitter account (@bankofengland) from its first tweet in July 2011 through July 2022
    \item \textbf{Public discourse}: All tweets containing the phrases "Bank of England" or "BoE" (case-insensitive) from 2007 through July 2022
\end{enumerate}

While "BoE" occasionally refers to entities unrelated to the Bank of England (e.g., "Board of Education"), we implemented rigorous cleaning procedures to minimise irrelevant matches. As acknowledged in prior research \citep{ehrmann2022socialmedia}, a small percentage of misclassified tweets inevitably remains—an inherent limitation of large-scale text scraping. The final dataset comprises approximately 3.13 million Bank of England-related tweets, with 9,810 tweets originating from the official BoE account between July 2011 and July 2022.

It is essential to note, as \cite{ehrmann2022socialmedia} emphasise, that Twitter users do not constitute a representative sample of the general population. This necessitates caution when extrapolating findings to broader public opinion. Nevertheless, the dataset provides valuable insights into how a significant segment of digitally active citizens encounters and engages with central bank communications.

\subsection{Descriptive Statistics}

The comprehensive dataset includes 3,126,016 tweets from 719,310 unique users spanning January 2007 through July 2022. Of these, 3,096,749 tweets (99.1\%) specifically mention the Bank of England. The data reveals substantial public engagement, with total interactions reaching 11,182,248 (averaging 3.58 engagements per tweet). This engagement comprises 939,753 replies, 3,047,777 retweets, 6,877,970 likes, and 316,748 quote tweets. The median engagement of zero, despite the 3.58 average, reflects the highly skewed distribution characteristic of social media interactions.

\begin{figure}[H]
  
    \includegraphics[width=0.9\textwidth, page=1]{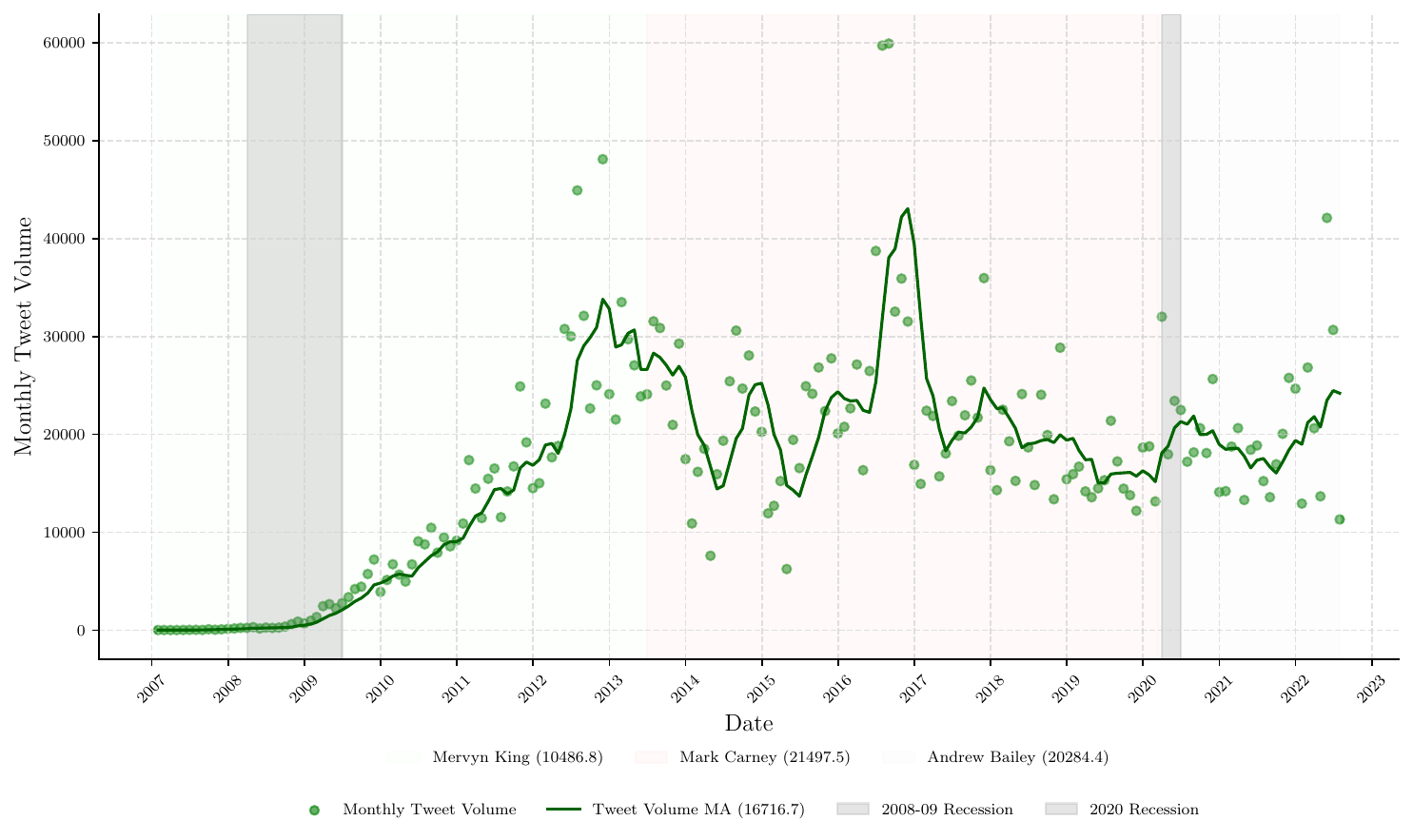}%
    \caption{Tweets mentioning "Bank of England" or "BoE" from 2007 through 2022}
      \label{f:tweetsoverview}
  \end{figure}

\subsubsection{Temporal Evolution of BoE-Related Twitter Activity}

The data reveals distinctive patterns in BoE-related Twitter activity over time. Figure 1 illustrates the monthly volume of tweets mentioning "Bank of England" or "BoE" from 2007 through 2022. Three distinct phases emerge:

\begin{enumerate}
    \item \textbf{Growth phase (2007-2013)}: Tweet volume expanded steadily, aligned with Twitter's overall user growth. Monthly references increased from mere dozens to consistently exceeding 10,000 by 2013.
    \item \textbf{Stabilization phase (2014-2016)}: Volume stabilized around a baseline of approximately 20,000 monthly tweets, with periodic spikes corresponding to significant economic events.
    \item \textbf{Volatility phase (2017-2022)}: While maintaining the baseline established in the previous phase, this period exhibits pronounced volatility, with dramatic spikes and troughs reflecting response to economic and political developments.
\end{enumerate}

The most remarkable spike occurred in August 2016, when Brexit-related news drove the monthly count to nearly 60,000 tweets (specifically 59,915), making 2016 the year with the highest volume at 388,584 tweets. Additional significant surges occurred during 2012-2013 (coinciding with unconventional monetary policy implementations), 2017 (Brexit negotiations), and most recently in May 2022, when mentions increased to over 42,000 amid heightened public concerns regarding inflation and cost-of-living debates.

\subsubsection{BoE's Official Twitter Activity and Language Distribution Patterns}

\begin{figure}[H]
    \includegraphics[width=0.85\textwidth, page=2]{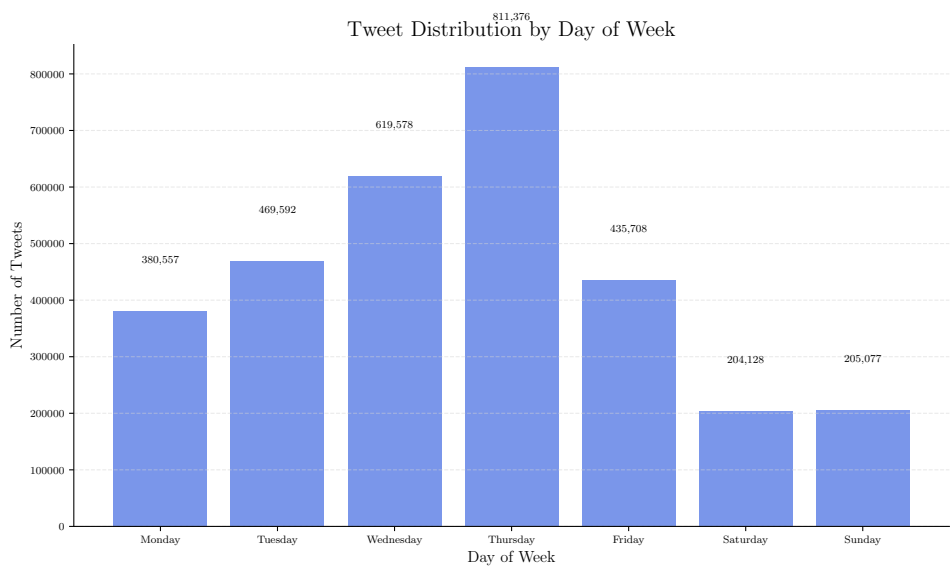}
      \caption{Tweet Distributions by Days}
      \label{f:days}
  \end{figure}

Examining weekly distribution patterns (Figure \ref{f:days}) reveals a pronounced concentration of BoE-related discourse during weekdays, particularly on Thursdays (26.0\% of all tweets, 811,376) and Wednesdays (19.8\%, 619,578). Weekend activity is substantially lower, with Saturday and Sunday accounting for just 6.5\% (204,128) and 6.6\% (205,077) of tweets respectively. This pattern aligns with the BoE's operational schedule and the timing of key policy announcements, which typically occur on weekdays—particularly Thursdays when the Monetary Policy Committee often announces decisions.

The vast majority of tweets in the dataset (97.5\%, 3,046,865) are in English, with small proportions in German (0.4\%, 11,962), Indonesian (0.3\%, 10,336), Dutch (0.3\%, 9,469), and Italian (0.2\%, 6,765). This linguistic concentration reflects both the BoE's primary operational jurisdiction and the international significance of its policy decisions for global financial markets.

\begin{figure}[H]
    \includegraphics[width=0.9\textwidth, page=3]{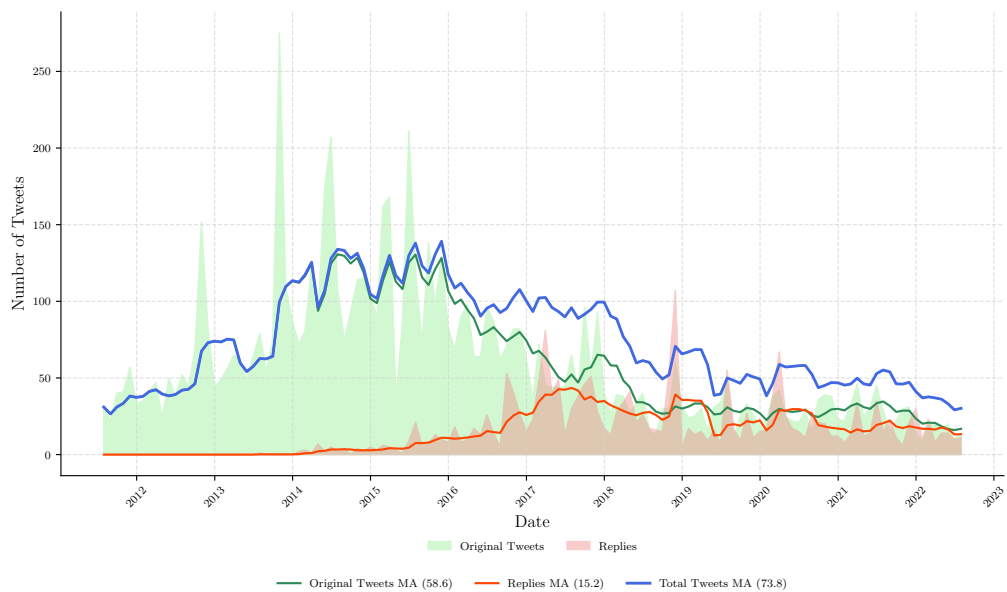}
      \caption{The BoE's official Twitter activity from 2011 through 2022}
      \label{f:officialtweets}
  \end{figure}

Figure-\ref{f:officialtweets} focuses exclusively on the Bank's own tweet volume and illustrates how its usage patterns evolved over the 11-year period. The data reveals distinct phases of activity with several notable peaks, including a maximum of 275 tweets in October 2013. Overall, the BoE posted an average of 73.8 tweets per month, with substantial variation in posting frequency across the period.

A crucial observation from Figure-\ref{f:officialtweets} is the shifting composition of the BoE's Twitter activity. The visualization distinguishes between original tweets (green color) and replies (peach color), revealing an increasing proportion of replies over time. While original tweets dominated until approximately 2015-2016, the reply component has grown significantly since then, representing a shift toward more interactive communication. Though replies still constitute a relatively small fraction of the Bank's total activity (2,021 reply tweets compared to 7,789 original tweets over the entire period), this interactive element reflects a significant transformation in central banking communication philosophy.

This evolution from unidirectional pronouncements to increasingly conversational engagement directly demonstrates what \cite{haldane2017little} describes as the shift from "communication means mouths" to "conversation means ears as much as mouths." Historically, senior central bank officials were reluctant to engage in public dialogue—for instance, Deputy Governor Harvey (1934) expressed "nervous[ness] at the thought of publication" and considered it "dangerous" to explain policy \citep{issing2019long}. The modern BoE Twitter presence represents a stark departure from that conservative stance, embodying the broader trend toward transparency in central banking.

\subsubsection{Hourly Engagement Patterns in Bank of England Twitter Activity}

Our analysis of Bank of England Twitter engagement reveals significant patterns in both the temporal distribution of engagement and the exceptional performance of certain tweets. The Bank's hourly engagement pattern shows distinct peaks and valleys, with the highest average engagement occurring at 22:00 UTC/GMT (128.0) and 06:00 UTC/GMT (106.0), despite these hours having relatively few tweets. This contrasts sharply with the most active posting time at 09:00, which accounts for 1,453 tweets (about 15\% of all tweets) but yields only moderate engagement (21.6 on average).

\begin{figure}[H]
    \includegraphics[width=0.8\textwidth, page=4]{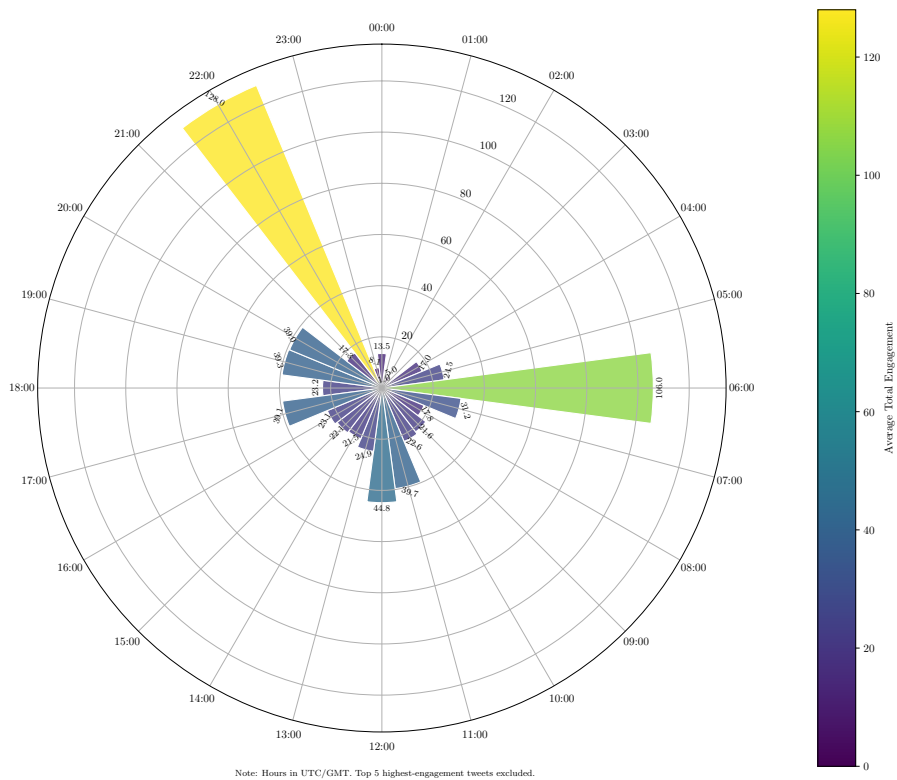}
      \caption{24-hour distribution of Bank of England tweet engagement}
      \label{f:24hourtweet}
  \end{figure}

  The significant disparity between peak posting hours and peak engagement hours suggests potential opportunities for the Bank to realign its Twitter communication strategy with periods of higher audience receptivity. Morning (6:00-11:59) and afternoon (12:00-17:59) periods account for 96.2\% of all tweets, with relatively comparable engagement metrics (means of 26.0 and 28.0 respectively), while evening tweets (18:00-23:59), though infrequent at just 3.6\% of total volume, show slightly higher average engagement (28.8).
  
The five most engaging tweets\footnote{To see the full list of tweets and their engagement metrics, please refer to the appendix} in the Bank of England's dataset dramatically outperform average metrics, with engagement levels ranging from 161.6 to 1,304.2 times the average. Notably, four of these five tweets (80\%) focus on a single topic: the introduction and release of the new \pounds50 note featuring Alan Turing. The top tweet ("\textit{\texttt{Introducing the face of the new \pounds50 note - the father of computer science, Alan Turing}}"), garnered extraordinary engagement with 35,189 total interactions, including 22,775 likes and 8,875 retweets (as of July 2022). This tweet alone achieved engagement 1,304.2 times higher than the average Bank tweet, demonstrating the exceptional public interest in currency design and commemorative aspects of central banking. The consistent theme across these high-performing tweets suggests that banknote announcements, particularly those celebrating notable historical figures like Turing, resonate strongly with public audiences which reflecting the continued cultural importance of cash, even in an increasingly cashless society.

The polar clock visualization (Figure \ref{f:24hourtweet}) effectively depicts the 24-hour distribution of Bank of England tweet engagement after excluding the five Turing £50 note outliers. Three distinct engagement peaks emerge: the most pronounced at 22:00 (128.0 average, bright yellow spoke), followed by 06:00 (106.0 average, green spoke), and a moderate midday band (11:00-12:00, blue spokes, averaging 39.7-44.8). In stark contrast, overnight hours (23:00-05:00) form an engagement desert with minimal activity (averages between 1.0-17.0), represented by short, dark purple spokes. This visualisation reveals a critical misalignment in the Bank's Twitter strategy—while tweet volume concentrates during business hours (peaking at 9:00 with 1,453 tweets), engagement peaks during early morning and late evening when posting is minimal. The substantial variation in hourly engagement metrics (coefficients ranging from 0.606 to 4.816) suggests significant opportunities for the Bank to optimise its posting schedule to better align with periods of audience receptivity.

\subsection{Engagement Metrics and Evolution}

\begin{figure}[H]
    \centering
    \includegraphics[width=0.8\textwidth, page=5]{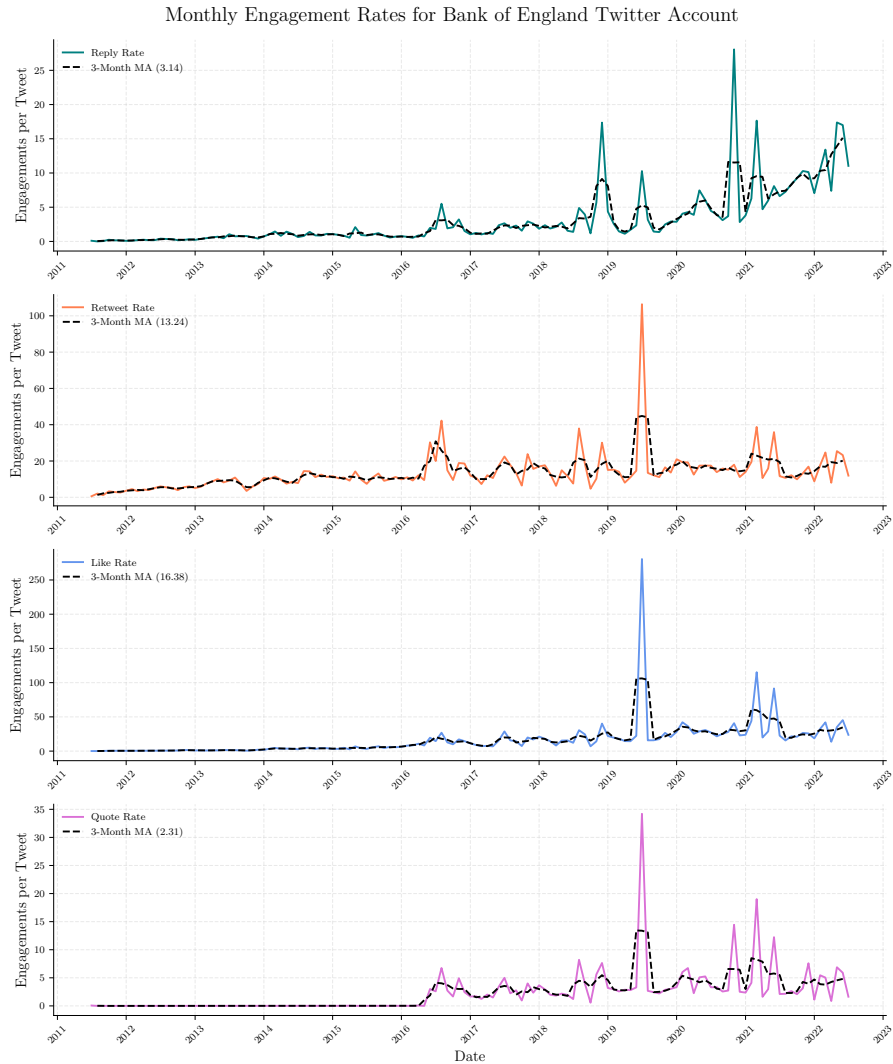}
    \caption{Monthly Engagement Rates of Individual Metrics}
    \label{fig:monthlyengagement}
\end{figure}

Figure \ref{fig:monthlyengagement} presents the monthly engagement metrics—replies, retweets, likes, and quote tweets—for the Bank of England's official Twitter account from July 2011 to July 2022. Each panel shows both raw engagement rates (colored line) and a three-month moving average (dotted black line), revealing distinct temporal patterns across the different engagement types.

For reply rates, we observe minimal activity (below 1 per tweet) until 2017 although the Bank started tweeting in 2011, followed by a gradual increase with a pronounced spike in November 2020 (28.06 replies per tweet) coinciding with a Transgender Awareness Week post that generated significant debate. Since 2020, reply rates have stabilised between 5-15 replies per tweet, indicating increased audience interaction in recent years. Retweet rates exhibit greater volatility, with the most substantial spike in July 2019 (106.39 retweets per tweet) during the Alan Turing £50 note announcement. The overall trajectory shows gradual growth from approximately 0.6 retweets per tweet in 2011 to a baseline of 15-20 by 2022.

Like rates demonstrate the most pronounced growth pattern, evolving from near-zero in 2011 to consistently exceeding 20 likes per tweet post-2019. The highest value (280.60 likes per tweet) occurred in July 2019 with the Turing announcement. Quote tweet rates remained at zero until Twitter introduced this feature in 2015, subsequently stabilizing at 2-5 quotes per tweet with a notable peak (34.19 quotes per tweet) during the July 2019 Turing announcement.

\begin{figure}[H]
        \centering
        \includegraphics[width=0.9\textwidth, page=6]{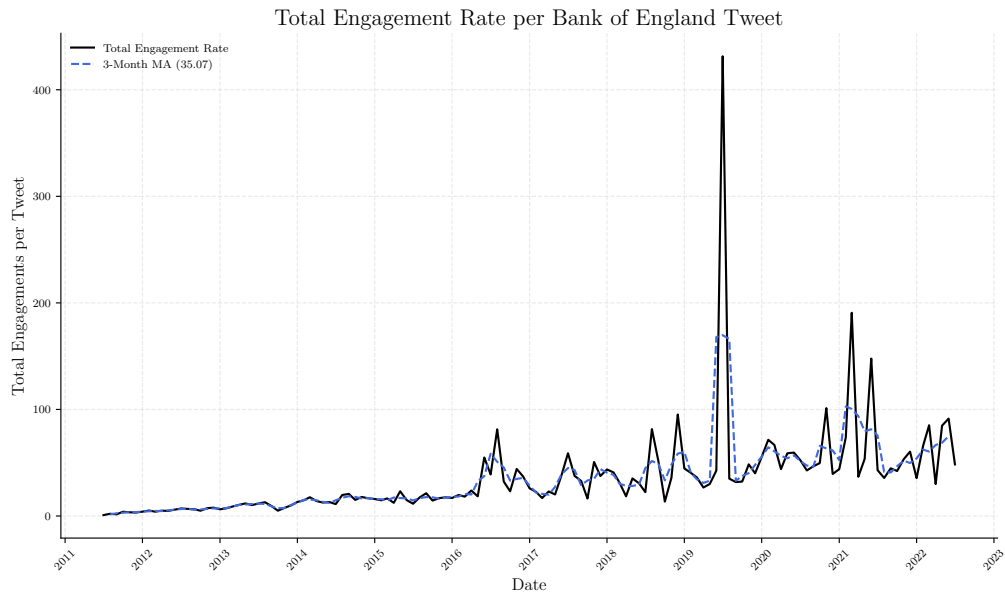}
        \caption{Total Engagement Rate}
        \label{fig:totalengagement}
    \end{figure}

Figure \ref{fig:totalengagement} aggregates these metrics into a total engagement rate, illustrating a fundamental shift from minimal engagement (under 5 interactions per tweet) before 2014 to substantially higher and more volatile engagement from 2016 onward. The three-month moving average indicates an exceptional spike in mid-2019 (exceeding 400 total engagements per tweet) followed by sustained higher engagement levels typically ranging from 40-100 engagements per tweet during 2020-2022.

Our longitudinal analysis spanning 133 months reveals considerable variation in engagement metrics. Average monthly engagement rates were 3.14 replies, 13.24 retweets, 16.38 likes, and 2.31 quotes per tweet, resulting in a total average engagement of 35.07 per tweet. These averages mask substantial temporal variation, with minimum values near zero in the early period (2011-2012) and maxima occurring predominantly in July 2019 (106.39 retweets, 280.60 likes, 34.19 quotes, and 431.46 total engagement), coinciding with the Turing £50 note announcement. Reply rates peaked later, reaching 28.06 in November 2020.

Annual summary statistics further confirm this evolution in engagement patterns. In 2011, the average tweet received just 0.57 total engagements (dominated by 0.13 retweets, with minimal likes at 0.01 per tweet). By 2022, this had increased to 14.21 engagements per tweet—representing a 25-fold increase. This growth accelerated notably during three periods: 2016 (1.90 engagements per tweet, coinciding with Brexit developments), 2018 (5.31 engagements per tweet), and 2019 (7.88 engagements per tweet, driven by the Turing announcement).

The composition of engagement has shifted significantly over time. In the early period (2011-2015), retweets constituted the primary form of engagement. By 2016, likes began approaching parity with retweets (0.87 vs. 0.79 per tweet). From 2017 onward, likes became increasingly dominant, reaching 10.20 per tweet in 2022 compared to 2.69 retweets, 1.00 replies, and 0.32 quotes per tweet. This compositional shift reflects both platform-specific changes in user behavior and evolving public interaction with central bank content.

The growth trajectory and engagement composition changes align with three distinct phases in the Bank's Twitter presence: an initial low-engagement period (2011-2013), a growth phase coinciding with increased Twitter adoption and heightened interest in monetary policy following Brexit (2014-2017), and a maturity phase (2018-2022) characterized by higher baseline engagement and occasional pronounced spikes during significant announcements. This evolution reflects both the Bank's developing social media strategy and changing user behavior on the platform.

\subsubsection{Content Analysis and High-Impact Tweets}

Analysis of the most engaged tweets reveals distinct patterns in public discourse surrounding the Bank of England. Table \ref{tab:top_tweets} presents the top ten  tweets about the Bank of England in our dataset, revealing two dominant themes: the Alan Turing £50 note announcement and Brexit's economic impact. The highest engagement (91,321 total interactions) was generated by the BBC Breaking News account announcing Turing as the face of the new banknote, significantly outperforming the official BoE announcement of the same news (35,222 engagements). This disparity illustrates how news intermediaries often generate greater public response than primary institutional sources.

\begin{table}[ht]
\caption{Top 10 Most Engaged Tweets About the Bank of England}
\label{tab:top_tweets}
\centering
\begin{tabular}{clccccc}
\toprule
\textbf{Rank} & \textbf{Username} & \textbf{Date} & \textbf{Replies} & \textbf{Retweets} & \textbf{Likes} & \textbf{Total} \\
\midrule
1 & BBCBreaking & 2019-07-15 & 1,559 & 16,547 & 68,905 & 91,321 \\
2 & Tim\_Burgess & 2022-05-16 & 983 & 13,220 & 47,516 & 62,593 \\
3 & nayibbukele & 2021-11-27 & 1,921 & 5,849 & 34,802 & 43,381 \\
4 & campbellclaret & 2022-05-15 & 1,189 & 5,051 & 31,105 & 37,569 \\
5 & DummiesEconomy & 2021-04-03 & 200 & 4,894 & 30,937 & 36,450 \\
6 & bankofengland & 2019-07-15 & 623 & 8,887 & 22,795 & 35,222 \\
7 & MartinSLewis & 2022-07-05 & 1,759 & 4,422 & 28,190 & 34,857 \\
8 & EdConwaySky & 2022-02-03 & 1,287 & 8,193 & 14,724 & 29,345 \\
9 & jonsnowC4 & 2017-06-20 & 766 & 13,723 & 12,786 & 28,654 \\
10 & EmmaKennedy & 2022-05-17 & 375 & 4,450 & 21,175 & 26,159 \\
\bottomrule
\end{tabular}
\end{table}

Notably, four of the ten highest-engagement tweets focus on Brexit's economic impact, including posts from \texttt{@Tim\_Burgess} highlighting "\textit{Brexit is costing the UK £444 million a week}" (62,593 interactions) and \texttt{@EmmaKennedy} noting similar costs. This pattern demonstrates how the BoE's economic assessments become focal points in broader political debates, often generating significantly more engagement when amplified through non-institutional voices.

The hashtag analysis presented in Figure \ref{f:tophashtags} provides insight into the thematic focus of BoE-related discourse. The most frequently used hashtags include \texttt{\#boe} (147,507 occurrences), \texttt{\#forex} (68,963), \texttt{\#bankofengland} (55,656), \texttt{\#brexit} (37,764), and \texttt{\#news} (36,414). The prominence of financial market terms (\texttt{\#forex}, \texttt{\#fx}, \texttt{\#trading}, \texttt{\#gbp}, \texttt{\#gbpusd}) alongside policy-related terms (\texttt{\#inflation}, \texttt{\#interestrates}) reflects how the BoE's communications are situated at the intersection of financial markets, economic policy, and broader public discourse. The significant presence of \texttt{\#brexit} as the fourth most common hashtag underscores how major political events reshape central bank communications.

\begin{figure}[H]
    \centering
  
    \subcaptionbox{Top Hashtags \label{f:tophashtags}}{%
        \includegraphics[width=0.48\textwidth, page=7]{\pdf}
    }\hfill
    \subcaptionbox{Tweet Word Cloud \label{f:tweetwordcloud}}{%
        \includegraphics[width=0.48\textwidth, page=8]{\pdf}
    }
    
    \caption{Tops Hashtag and Word Cloud}
    \label{fig:hastagcloud}
\end{figure}



The word cloud visualization in Figure \ref{f:tweetwordcloud} further illustrates the dominant themes in discourse surrounding the Bank of England. Terms like "interest," "rate," "Brexit," "inflation," and "governor" feature prominently, reflecting the primary policy concerns and institutional leadership that drive public discussion. The prominence of "Carney" indicates the personification of central banking leadership in public discourse, while economic terms like "pound," "sterling," and "QE" highlight the focus on monetary policy instruments.

\subsubsection{Influential Accounts and Network Dynamics}

Our analysis reveals a significant disparity between posting frequency and engagement impact among accounts discussing the Bank of England. Table \ref{tab:influential_accounts} presents the most influential accounts in the dataset, measured by total engagement generated.

\begin{table}[ht]
\caption{Most Influential Accounts in Bank of England Twitter Discourse}
\label{tab:influential_accounts}
\centering
\begin{tabular}{clrrr}
\toprule
\textbf{Rank} & \textbf{Username} & \textbf{Total Tweets} & \textbf{Total Engagement} & \textbf{Avg. Engagement} \\
\midrule
1 & RichardJMurphy & 755 & 374,964 & 496.6 \\
2 & bankofengland & 9,952 & 321,164 & 32.3 \\
3 & BBCBreaking & 140 & 147,275 & 1,052.0 \\
4 & business & 2,253 & 109,901 & 48.8 \\
5 & johnredwood & 88 & 84,180 & 956.6 \\
6 & BTC\_Archive & 22 & 79,366 & 3,607.5 \\
7 & EdConwaySky & 1,345 & 71,531 & 53.2 \\
8 & Peston & 249 & 71,381 & 286.7 \\
9 & SkyNews & 950 & 70,436 & 74.1 \\
10 & nayibbukele & 2 & 68,821 & 34,410.5 \\
\bottomrule
\end{tabular}
\end{table}

This data reveals several intriguing patterns. First, while the official \texttt{@bankofengland} account has been highly active (9,952 tweets) and ranks second in total engagement (321,164), its average engagement per tweet (32.3) is dramatically lower than many non-official commentators. For instance, \texttt{@RichardJMurphy} achieved greater total impact (374,964 engagement across 755 tweets) with less than 8\% of the BoE's posting volume, resulting in an average engagement per tweet (496.6) more than 15 times higher than the official account.

The highest average engagement belongs to \texttt{@nayibbukele} with an extraordinary 34,410.5 engagements per tweet, though this is based on only two tweets in the dataset. Other accounts like \texttt{@BBCBreaking} (1,052.0 average engagement) and \texttt{@johnredwood} (956.6 average engagement) consistently achieve high engagement levels while maintaining selective posting practices about BoE-related topics.

In contrast, the most prolific posters beyond the BoE itself include market-focused accounts like \texttt{BlackCentaurFX} (9,346 tweets), \texttt{notayesmansecon} (8,224 tweets), and \texttt{FXStreetNews} (6,605 tweets) – yet none of these appear among the top 10 for influence, suggesting quantity does not necessarily translate to impact.

This discrepancy underscores an important distinction in digital central bank communication: official central bank accounts typically maintain a consistent, high-volume communication strategy focused on transparency and information dissemination, while non-official commentators can generate dramatically higher engagement rates by selectively amplifying and contextualizing central bank communications – particularly when these intersect with politically contentious issues like Brexit or when they frame technical financial information for broader audiences. The BoE's communications are thus filtered, amplified, and recontextualized within broader public discourse, often reaching their widest audience through intermediaries rather than directly.

The data further suggests that while central banks can generate exceptionally high engagement with cultural or human interest content (such as the Turing banknote announcement), their core monetary policy and financial stability communications typically achieve more modest engagement levels, highlighting the challenge of making technical economic content accessible and engaging to broader public audiences.

\subsubsection{Summary and Implications}

\begin{table}[H]
    \caption{Official BoE Twitter Engagement Statistics (July 2011–July 2022)}
    \begin{tabular*}{\textwidth}{p{4cm}@{\extracolsep\fill}ccc}
    \toprule
        \textbf{Metric} & \textbf{Mean} & \textbf{Min} & \textbf{Max} \\
    \midrule
    Tweets (count) & 9,810 & -- & -- \\
    Replies per tweet & 2.45 & 0.00 & 28.06 \\
    Retweets per tweet & 13.12 & 0.58 & 106.39 \\
    Likes per tweet & 14.92 & 0.05 & 280.60 \\
    Quotes per tweet & 2.06 & 0.00 & 34.19 \\
    Total engagement & 32.55 & 0.81 & 431.46 \\
    \bottomrule
    \end{tabular*}
    \note[Note]{This table presents the distribution of engagement metrics for official BoE tweets over the study period, showing the mean, minimum, and maximum values for each engagement type.}
    \label{t:engagement_stats}
\end{table}

Over the entire study period, the BoE's official posts receive, on average, about 2.45 replies, 13.12 retweets, 14.92 likes, and 2.06 quotes, for a total of 32.55 engagements per tweet (Table \ref{t:engagement_stats}). Individual monthly extremes can reach much higher levels, as seen with the Turing \pounds50 note reveal in mid-2019, which produced engagement levels more than 1,000 times the daily average of the earliest years in the sample.

Our analysis demonstrates several key findings about the Bank of England's Twitter engagement. First, there exists a significant misalignment between posting patterns and engagement opportunities, with the Bank concentrating most posts during business hours while audience engagement peaks in early morning and evening hours. Second, content related to currency design and wider socioeconomic issues like Brexit generates substantially higher engagement than core monetary policy announcements. Third, engagement has evolved dramatically over time, with a 1,000-fold increase in average engagement rates and a shift from retweet-dominated to like-dominated interaction patterns. Finally, the Bank's communications are frequently amplified and recontextualized by non-official accounts, which often generate higher per-tweet engagement than the Bank's official account.

These findings provide the foundation for our subsequent investigation of the determinants of public engagement with central bank communications, offering potential strategic insights for optimising the timing, content, and approach of central bank social media communications.

\section{Methodology}\label{sec:methodology}

\subsection{Model Specifications}

To investigate engagement with the Bank of England (BoE) on Twitter, we employ two complementary empirical approaches. The first examines whether the volume of BoE tweets drives public engagement, while the second explores specific tweet characteristics that influence engagement levels.

\subsubsection{Model 1: Elasticity of Response to Tweet Volume}

Following \cite{gorodnichenko2024socialmedia}, we first estimate the elasticity of public response to the number of official BoE tweets. The model specification is:

\begin{equation} \label{eq:model1}
    \ln (\text{Reaction})_{j,w} = \alpha + \beta \ln (\text{Number of Tweets})_{j,w} + \epsilon_{j,w}
\end{equation}

Where:
\begin{itemize}
    \item $\ln (\text{Reaction})_{j,w}$ represents the natural logarithm of one plus the size of reaction $j$ in week $w$. We consider four distinct reaction dimensions: likes, replies, retweets, and quote tweets.
    \item $\ln (\text{Number of Tweets})_{j,w}$ is the natural logarithm of one plus the number of official BoE tweets in week $w$.
    \item $\epsilon_{j,w}$ is the error term.
\end{itemize}

This (Model-\ref{eq:model1}) specification estimates the percentage change in public engagement associated with a percentage change in BoE tweet frequency. We estimate this equation separately for each month using Ordinary Least Squares with heteroskedasticity-robust standard errors, generating a monthly time series of elasticities for each engagement dimension. This approach differs from \cite{gorodnichenko2024socialmedia} in its inclusion of quote tweets as an additional reaction dimension, providing a more comprehensive assessment of engagement patterns. The resulting elasticity series reveals how the relationship between tweet volume and public response evolves over time and across different engagement types.

\subsubsection{Model 2: Determinants of Tweet-Level Engagement}

The second model explores the specific characteristics of tweets that drive engagement. We adapt Model 2 from \cite{gorodnichenko2024socialmedia} to better suit the BoE context:

\begin{equation} \label{eq:model2}
\text{Reaction}_{i,d} = \alpha + \beta_1 \text{MPC}_{d} + \beta_2 \text{Characteristics}_{i} + \epsilon_{i,d} \end{equation}

Where:
\begin{itemize}
\item $\text{Reaction}_{i,d}$ represents the count of a specific engagement type (likes, replies, retweets, or quote tweets) for tweet $i$ on date $d$.
\item  $\text{MPC}_{d}$ is a dummy variable equal to one if the Monetary Policy Committee makes an announcement on date $d$, and zero otherwise.
\item $\text{Characteristics}_{i}$ is a vector of tweet characteristics including content features (presence of links, hashtags, media types) and linguistic attributes (complexity).
\item $\epsilon_{i,d}$ is the error term.
\end{itemize}

Our specification diverges from \cite{gorodnichenko2024socialmedia} in several important ways. First, we employ separate dummy variables for different tweet types rather than a categorical variable, allowing for more nuanced effect size estimation. Second, we include a broader range of content characteristics, particularly distinguishing between different media types (photos, videos, GIFs) rather than treating all media as homogeneous. Third, we incorporate a measure of linguistic complexity (Flesch Reading Ease) to assess how readability affects engagement.

Since the dependent variables are mostly count data, we employ Poisson regression rather than OLS, which is more appropriate for non-negative integer outcomes with right-skewed distributions. This approach provides a more accurate estimation of effects when modeling count data.

\subsection{Data and Variables}

\subsubsection{Data Collection and Processing}

The dataset comprises all tweets from the official Bank of England Twitter account (@bankofengland) from its first tweet in July 2011 through July 2022. This encompasses 9,810 tweets in total, providing a comprehensive view of the BoE's social media communication strategy over more than a decade.

For Model-\ref{eq:model1}, we aggregate the data into weekly observations to capture short-term fluctuations in tweet volume and engagement, then estimate monthly elasticities to observe evolving patterns. For Model-\ref{eq:model2}, we conduct analysis at the individual tweet level to identify specific features that drive engagement.

\subsubsection{Variable Definitions}

\textbf{Dependent Variables:}

\begin{itemize}
    \item \textbf{Likes}: Count of likes received by each tweet
    \item \textbf{Replies}: Count of direct replies to each tweet
    \item \textbf{Retweets}: Count of retweets (shares without additional comment)
    \item \textbf{Quote Tweets}: Count of retweets with additional commentary
\end{itemize}

\noindent \textbf{Independent Variables:}

\begin{itemize}
    \item \textbf{MPC Announcement (Is\_mpc)}: Binary indicator for tweets posted on days when the Monetary Policy Committee makes policy announcements
    \item \textbf{Reply Status (isReply)}: Binary indicator for whether the tweet is a reply to another user
    \item \textbf{Link Inclusion (isLink)}: Binary indicator for tweets containing hyperlinks
    \item \textbf{Hashtag Inclusion (isHashtag)}: Binary indicator for tweets containing hashtags
    \item \textbf{Media Type}: Mutually exclusive binary indicators for different media types:
    \item \textbf{GIF}: Tweets containing animated GIF images
    \item \textbf{Photo}: Tweets containing static images
    \item \textbf{Video}: Tweets containing video content
    \item \textbf{Complexity}: Continuous measure of linguistic complexity using Flesch Reading Ease (higher values indicate greater readability)
\end{itemize}

Additional control variables (not shown in the main results but included in the full model) account for temporal patterns:
\begin{itemize}
    \item \textbf{Day of week (weekday dummies)}
    \item \textbf{Hour of day (hour dummies)}
    \item \textbf{Tweet sequence within a day (day\_num and day\_num\_sq)}
\end{itemize}

This comprehensive set of variables allows for a detailed examination of the factors that influence public engagement with central bank communications on social media.

\section{Results} \label{sec:results}

\subsection{Model 1: Elasticity of Response to Tweet Volume}

\begin{figure} [H]
        \centering
        \includegraphics[width=0.9\textwidth, page=9]{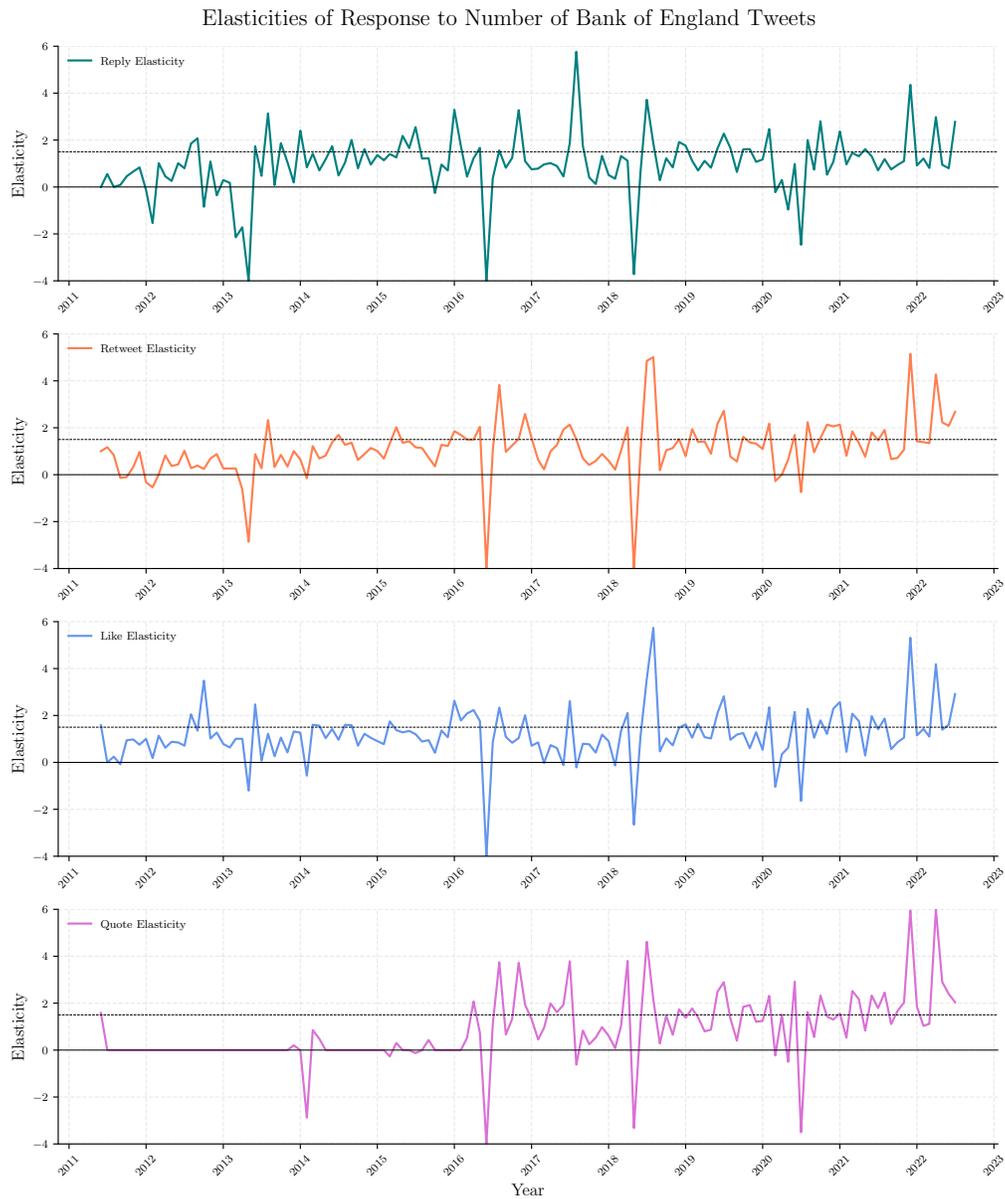}
        \caption{Elasticity of Individual Metrics}
        \label{fig:sub1x}
    \end{figure}

Figures \ref{fig:sub1x} and \ref{fig:sub2x} present the estimated elasticities of public engagement with respect to the Bank of England's tweet volume over the 2011-2022 period. These elasticity measures capture the percentage change in engagement associated with a one percent change in the Bank's monthly tweet frequency, providing insight into the efficiency of its communication strategy.

The elasticity estimates reveal distinct patterns across different engagement dimensions. Reply elasticity averages 0.984 (min: -4.000, max: 5.763), indicating that replies typically scale almost proportionally with tweet volume. The considerable volatility in these values suggests that audience willingness to engage in conversations with the Bank depends substantially on content relevance rather than posting frequency alone. Retweet elasticity exhibits a slightly higher mean of 1.088 (min: -4.000, max: 5.151), demonstrating marginally increasing returns to scale, with notable elasticity peaks coinciding with significant policy announcements.

Like elasticity displays the highest average responsiveness to tweet volume changes at 1.164 (min: -4.000, max: 5.738), consistent with the growing prominence of likes as the dominant engagement mechanism throughout the study period. Quote elasticity shows a distinctive temporal pattern with a mean of 0.837 (min: -4.000, max: 6.000), beginning at zero before quotes were introduced as a Twitter feature in 2015, then displaying increasingly volatile behaviour in subsequent years.

The consolidated total engagement elasticity averages 1.095 across the entire period, indicating slightly increasing returns to scale in overall public engagement with Bank of England tweets. However, this aggregate measure masks significant temporal variation characterized by three notable patterns:

\begin{figure} [H]
        \centering
        \includegraphics[width=0.9\textwidth, page=10]{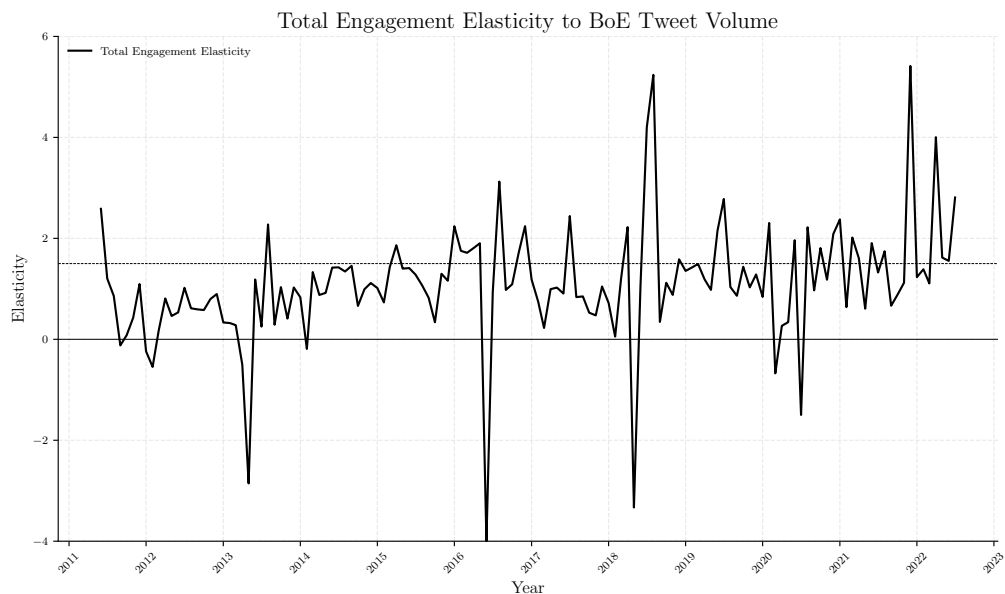}
        \caption{Elasticity of Total Engagement Over Time}
        \label{fig:sub2x}
    \end{figure}

First, the elasticity series exhibits periodic negative values, particularly during 2013, 2016, and early 2018, representing intervals when increased tweet frequency coincided with reduced engagement per tweet. These instances suggest diminishing returns or possible audience fatigue during periods of high-volume Bank communications. Second, pronounced positive spikes exceeding 5.0 typically correspond to months with relatively low tweet volumes but exceptionally high-impact content. The most dramatic example occurred in July 2019 with the Turing £50 note announcement, when the Bank achieved extraordinary engagement despite posting fewer tweets than in typical months.  Third, elasticity volatility has intensified noticeably since 2017, with more frequent and pronounced oscillations between positive and negative values. This increasing unpredictability suggests that as the social media landscape matures, audience response to central bank communications becomes more complex and content-dependent.

Unlike the Federal Reserve, which \cite{gorodnichenko2024socialmedia} found to have relatively stable elasticities mostly between 0 and 1.5, the Bank of England exhibits substantially greater volatility in its engagement elasticities. The most extreme values occurred in June 2016, coinciding with the Brexit referendum, when elasticities for quote tweets and replies reached -60 and -20 respectively, indicating that despite fewer tweets, each individual post generated dramatically higher engagement during this period of heightened economic uncertainty.

The pronounced co-movement across different engagement dimensions suggests common underlying factors driving overall engagement beyond mere posting frequency. This finding aligns with \cite{ehrmann2022socialmedia}, who emphasized that content characteristics significantly influence public response to central bank communications. The observed elasticity patterns indicate that for the Bank of England, content quality, timing, and relevance to current economic conditions appear to matter substantially more than posting volume for maximizing public engagement.

These results have important implications for central bank communication strategies. While modestly increasing tweet frequency may generate slightly more than proportional increases in engagement on average, the substantial temporal variation in elasticities suggests that a more refined approach focused on strategic timing and high-impact content would likely yield greater public engagement than simply increasing posting volume.

\subsection{Model 2: The Importance of Content Characteristics: Determinants of Tweet-Level Engagement}

Table \ref{t:engagement_determinants} presents the results from the Poisson regression model estimating the factors that influence engagement with BoE tweets. The coefficients can be interpreted by exponentiating them and subtracting one, yielding the percentage change in the expected count of the engagement metric associated with each feature.

\subsubsection{Monetary Policy Committee Announcements}

Tweets posted on MPC announcement days receive substantially higher engagement across all metrics. Specifically, compared to non-MPC days, these tweets receive:
\begin{itemize}
    \item 122\% more likes ($e^{0.797} - 1 = 1.22$)
    \item 270\% more retweets ($e^{1.307} - 1 = 2.70$)
    \item 123\% more replies ($e^{0.801} - 1 = 1.23$)
    \item 376\% more quote tweets ($e^{1.562} - 1 = 3.76$)
\end{itemize}

This substantial effect highlights the central importance of monetary policy announcements in driving public engagement with the BoE on social media. The particularly strong effect on retweets and quote tweets suggests that monetary policy news is especially likely to be shared and commented upon, extending the reach of these communications beyond the BoE's immediate followers.

\begin{table}[H]
\caption{Determinants of Tweet-Level Engagement}
\begin{adjustbox}{width=0.89\textwidth}
\begin{tabular*}{\textwidth}{p{3cm}@{\extracolsep\fill}cccc}
\toprule
& \multicolumn{4}{c}{Engagement Metrics}\\
\cmidrule{2-5}
& Likes & Retweets & Replies & Quote Tweets\\
\midrule
Constant & -0.084*** & 1.782*** & -0.723*** & -3.344***\\
& (0.027) & (0.025) & (0.069) & (0.102)\\
\midrule
MPC Announcement & 0.797*** & 1.307*** & 0.801*** & 1.562***\\
& (0.012) & (0.011) & (0.027) & (0.028)\\
\midrule
Reply Status & -1.130*** & -1.930*** & -0.585*** & -1.074***\\
& (0.013) & (0.017) & (0.023) & (0.034)\\
\midrule
Link Inclusion & 0.696*** & 0.345*** & 0.012 & 0.647***\\
& (0.007) & (0.007) & (0.016) & (0.019)\\
\midrule
Hashtag Inclusion & 0.419*** & 0.264*** & 0.089*** & 0.437***\\
& (0.006) & (0.006) & (0.015) & (0.017)\\
\midrule
GIF & 1.365*** & 1.034*** & 1.454*** & 2.302***\\
& (0.018) & (0.016) & (0.040) & (0.042)\\
\midrule
Photo & 2.268*** & 1.159*** & 1.778*** & 2.740***\\
& (0.007) & (0.006) & (0.016) & (0.022)\\
\midrule
Video & 2.923*** & 1.838*** & 2.204*** & 3.575***\\
& (0.011) & (0.012) & (0.029) & (0.029)\\
\midrule
Complexity & 0.016*** & 0.006*** & 0.013*** & 0.018***\\
& (0.000) & (0.000) & (0.000) & (0.000)\\
\bottomrule
\end{tabular*}
\end{adjustbox}
\note{Note: Standard errors in parentheses. * p<0.1, ** p<0.05, *** p<0.01.}
\label{t:engagement_determinants}
\end{table}

\subsubsection{Tweet Type and Content Characteristics}

\textbf{Reply Status}: Tweets that are replies to other users receive significantly less engagement than original tweets across all metrics:

\begin{itemize}
    \item 68\% fewer likes ($e^{-1.13} - 1 = -0.68$)
    \item 85\% fewer retweets ($e^{-1.93} - 1 = -0.85$)
    \item 44\% fewer replies ($e^{-0.585} - 1 = -0.44$)
    \item 66\% fewer quote tweets ($e^{-1.074} - 1 = -0.66$)
\end{itemize}

This substantial reduction in visibility for reply tweets aligns with Twitter's algorithmic design, which gives less prominence to replies in followers' timelines.

\textbf{Links and Hashtags}: Both links and hashtags are associated with increased engagement:
\begin{itemize}
    \item Tweets with links receive 101\% more likes and 41\% more retweets
    \item Tweets with hashtags receive 52\% more likes and 30\% more retweets
\end{itemize}

Interestingly, links show no significant effect on replies (p=0.434), suggesting that while they encourage sharing, they may not stimulate conversation to the same degree.

\subsubsection{Media Content}

The inclusion of media substantially boosts engagement, with different media types showing varying effects:

\textbf{Photos} increase engagement across all metrics:
\begin{itemize}
    \item 126\% more likes
    \item 219\% more retweets
    \item 120\% more replies
    \item 380\% more quote tweets
\end{itemize}

\textbf{GIFs} demonstrate similar positive effects:
\begin{itemize}
    \item 190\% more likes
    \item 81\% more retweets
    \item 220\% more replies
    \item 800\% more quote tweets
\end{itemize}

In summary, the results from the gif and photos suggest that these are another powerful tool for central bank communication that merits special attention which is infographics. A well-designed infographics that present complex economic concepts through simple visual elements can significantly enhance public understanding of central bank policies. The combination of visual appeal and simplified messaging creates an accessible entry point for non-expert audiences, potentially generating engagement levels comparable to other media types. Central banks could leverage infographics to break down complex policy decisions into digestible components, making their communications more inclusive and effective across diverse audience segments.

\textbf{Videos} show the strongest impact of all media types:

\begin{itemize}
    \item 1700\% more likes
    \item 500\% more retweets
    \item 700\% more replies
    \item 3300\% more quote tweets
\end{itemize}

The exceptionally strong effect of video content suggests that this media format is particularly effective for central bank communications. However, as noted in the methodology, these coefficients should not necessarily be interpreted as causal effects, as videos are typically reserved for the most important announcements (such as new banknote designs), which might generate high engagement regardless of format.

\subsubsection{Linguistic Complexity}

The complexity variable shows a positive coefficient across all engagement metrics, indicating that more readable content (higher Flesch Reading Ease scores) generates greater engagement. Specifically, a one-point increase in readability is associated with:

\begin{itemize}
    \item 1.6\% more likes
    \item 0.6\% more retweets
    \item 1.3\% more replies
    \item 1.8\% more quote tweets
\end{itemize}

While these effects appear modest for a single-point change, they become substantial when considering the typical range of readability scores. This finding provides empirical support for the BoE's efforts to simplify its communications, suggesting that clearer, more accessible language yields tangible benefits in terms of public engagement.

\subsection{Implications of Findings}

These results provide several insights into effective central bank communication strategies on social media:

\begin{itemize}
    \item \textbf{Content matters more than frequency}: The elasticity analysis demonstrates that for the BoE, the quality and relevance of content drive engagement more than posting volume. This contrasts with findings for the Federal Reserve, suggesting potential differences in audience composition or expectations.

    \item \textbf{Monetary policy events create engagement opportunities}: The strong positive effect of MPC announcement days highlights the importance of these key policy moments for public engagement. The BoE could leverage this heightened attention by ensuring particularly clear and accessible communication during these periods.

    \item \textbf{Media enhances engagement}: The substantial positive effects of photos, GIFs, and especially videos underscores the importance of visual communication in making central bank messages more engaging. The exceptionally strong effect of video content suggests this medium may be particularly valuable for communicating complex economic concepts.

    \item \textbf{Simplicity enhances reach}: The positive relationship between readability and engagement supports the BoE's efforts to simplify its communications. This finding aligns with \cite{haldane2018little} advocacy for clearer central bank language and suggests that accessibility does not come at the cost of engagement—it enhances it.

    \item \textbf{Reply tweets face visibility challenges}: The significantly lower engagement with reply tweets highlights a potential limitation in using Twitter for two-way communication. To overcome this, the BoE might consider alternative approaches to direct engagement that maintain higher visibility.
\end{itemize}
These findings contribute to our understanding of how central banks can effectively communicate with the public in the digital age. They suggest that the BoE's strategy would benefit from continuing to focus on clear, accessible messaging with strategic use of media, particularly during key policy announcements.

\section{Discussion and Conclusion}\label{sec:discussion}

\paragraph{Theoretical Implications} The findings from both models contribute to theoretical discussions about central bank communication in several ways. First, they provide empirical support for \cite{haldane2018little} argument for a shift from traditional unidirectional communication ("communication means mouths") to more conversational, accessible engagement ("conversation means ears as much as mouths"). The strong positive relationship between readability and engagement substantiates that simpler language enhances public interaction with central bank messages.

Second, the results extend \cite{blinder2024communicationgeneralpublic} research on central bank communication with the general public by quantifying how different message characteristics affect engagement. The exceptional performance of media-rich content, particularly videos, aligns with evidence from \cite{bholat2019enhancing} that visual elements significantly improve comprehension and retention of central bank communications.

Third, the pronounced elasticity volatility for the BoE compared to the Federal Reserve \citep{gorodnichenko2024socialmedia} suggests important cross-institutional differences in how audiences engage with central bank communications. This finding cautions against universal prescriptions for central bank communication strategies and highlights the importance of institution-specific approaches tailored to particular audiences and contexts.

\paragraph{Communication Strategy Implications} The results suggest several practical implications for the BoE's social media strategy. First, the significant engagement boost on MPC announcement days represents an opportunity to capitalize on heightened public attention. The Bank could enhance its impact by concentrating particularly accessible, media-rich content on these days when the public is most receptive.

Second, the strong positive effects of media content, particularly videos, suggest the BoE should continue expanding its use of visual communication. While videos show the strongest effect, their production requires substantial resources. Photos and GIFs offer considerable engagement benefits with lower production costs, potentially offering better efficiency for routine communications.

Third, the finding that reply tweets receive substantially less engagement presents a challenge for two-way communication. This suggests the BoE might benefit from alternative approaches to dialogue, such as periodic Q\&A sessions or threading replies with original tweets to maintain visibility.

Fourth, the positive effect of readability on engagement provides empirical validation for the BoE's plain language initiatives. This finding suggests that continuing to simplify communications will not only improve public understanding, as advocated by \cite{hansen2018bank} and \cite{haldane2018little}, but will also expand the reach of the Bank's messages.

\paragraph{Relationship to Previous Literature} These findings both complement and extend previous research on central bank communication. The importance of media content aligns with \cite{masciandaro2024central_joes}, who found that announcements about new banknotes generate disproportionate engagement. Our results quantify this effect more precisely, showing that tweets with video content receive up to 18 times more likes than text-only tweets.

The elasticity patterns for the BoE differ notably from those found by \cite{gorodnichenko2024socialmedia} for the Federal Reserve, where elasticities were relatively stable around 1.0. This divergence suggests that BoE audiences may respond differently to communication patterns than Federal Reserve audiences, highlighting the importance of institution-specific research in central bank communication.

The finding that readability enhances engagement supports \cite{korhonen2024microblogging} conclusion that greater clarity leads to stronger engagement with ECB communications. It also provides empirical validation for \cite{lamla2022gentlemantweet} observation that the BoE stands out for its efforts to use clear and accessible language on Twitter.

The robust effect of MPC announcements on engagement aligns with \cite{hansen2018bank} finding that specific wording in BoE communications measurably affects financial markets. Together, these results underscore how both the timing and content of central bank communications influence public response.

\paragraph{Limitations and Future Research} Several limitations should be acknowledged. First, while the models identify correlations between tweet characteristics and engagement, they cannot definitively establish causality. For instance, the strong effect of video content may partly reflect that videos are reserved for inherently more engaging announcements. Second, as acknowledged by \cite{ehrmann2022socialmedia}, Twitter users do not constitute a representative sample of the general population. Therefore, these findings may not generalize to the broader public's engagement with central bank communications. Future research could address these limitations through experimental approaches that isolate causal effects of specific content features. Additionally, combining social media engagement data with surveys of public understanding would provide a more comprehensive assessment of communication effectiveness beyond mere engagement metrics. Cross-institutional comparisons of engagement patterns across different central banks would also be valuable for understanding whether the patterns observed for the BoE reflect institution-specific factors or broader trends in central bank communication.

\paragraph{Conclusion} This analysis of Bank of England Twitter (X) engagement reveals several key insights for central bank communication in the digital age. Unlike the Federal Reserve, where engagement responds proportionally to tweet frequency, the BoE's audience engagement depends more critically on content characteristics than volume. Monetary policy announcements significantly boost engagement, as does media-rich content and more readable language. These findings provide empirical support for the BoE's efforts to enhance accessibility and visual appeal in its communications. They suggest that effective central bank communication on social media requires not just regular posting but strategic content optimization focused on readability, visual elements, and capitalizing on moments of heightened public attention.

The results contribute to a growing literature on central bank communication with non-expert audiences, demonstrating that engagement metrics can provide valuable feedback on communication effectiveness. As central banks continue expanding their digital outreach, understanding these engagement dynamics will be increasingly important for fulfilling their public communication mandate.

\singlespacing
\bibliography{\bib}
\end{document}